\documentclass[aps,prb,twocolumn,showpacs]{revtex4-1}
\usepackage{graphicx}
\begin{document}

\title{An experimental and theoretical study \\
of Ni impurity centers in Ba$_{0.8}$Sr$_{0.2}$TiO$_3$}

\author{I. A. Sluchinskaya}
\email[]{irinasluch@gmail.com}
\affiliation{Moscow State University, Moscow, 119991 Russia}
\author{A. I. Lebedev}
\affiliation{Moscow State University, Moscow, 119991 Russia}

\date{\today}

\begin{abstract}
The local environment and the charge state of a nickel impurity in cubic
Ba$_{0.8}$Sr$_{0.2}$TiO$_3$ are studied by XAFS spectroscopy. According to the
XANES data, the mean Ni charge state is $\sim$2.5+. An analysis of the EXAFS
spectra and their comparison with the results of first-principle calculations
of the defect geometry suggest that Ni$^{2+}$ ions are in a high-spin state at
the $B$~sites of the perovskite structure and the difference of the Ni$^{2+}$
and Ti$^{4+}$ charges is mainly compensated by distant oxygen vacancies. In
addition, a considerable amount of nickel in the sample is in a second phase
BaNiO$_{3-\delta}$. The measurements of the lattice parameter show a decrease
in the unit cell volume upon doping, which can indicate the existence of
a small amount of Ni$^{4+}$ ions at the $B$~sites.

\texttt{DOI: 10.1134/S106378341708025X}
\end{abstract}


\maketitle

\section{INTRODUCTION}

In recent years, ferroelectric oxides with the perovskite $AB$O$_3$ structure
attract much attention when discussing a possibility of designing various
solar energy converters based on these materials. In particular, the idea of
practical application of the bulk photovoltaic effect consisting in the
appearance of a photocurrent or very high photovoltages upon illumination
of homogeneous crystals without an inversion center (including
ferroelectrics)~\cite{Sturman-Fridkin} has been discussed since
1970s.~\cite{ApplPhysLett.25.233} However, since the quantum yield of this
effect was low, the idea was not used in practice. In recent years, the
interest to ferroelectric oxides was revitalized due to the appearance of
new ideas how to increase the efficiency of solar energy converters based on
the bulk photovoltaic effect.~\cite{ApplPhysLett.93.122904,NatureCommun.2.256}
The main disadvantage of ferroelectric oxides is their comparatively wide
energy gap ($\sim$3~eV), as a result of which they absorb only a small part
of solar radiation. To solve this problem, it has long been proposed to dope
the oxides with $3d$ elements~\cite{JInorgNuclChem.43.1499} that in some cases
can result in the formation of color centers.

Recent theoretical works showed that the substitution for Ti in PbTiO$_3$ with
an impurity having $d^8$ electron configuration (Ni, Pd, Pt) compensated by
an oxygen vacancy made it possible to decrease the energy gap of this
material.~\cite{JAmChemSoc.130.17409,PhysRevB.83.205115} The experimental
studies of SrTiO$_3$ doped with a number of $3d$ elements (Mn, Fe, Co, and
Ni)~\cite{JETPLett.89.457,BullRASPhys.74.1235,th1,th2,JAdvDielectrics.3.1350031,
PhysSolidState.56.449} showed that the doping of SrTiO$_3$ results in a
strong absorption of light in the visible region and this effect is the most
significant for Co and Ni impurities. However, since strontium titanate is
not a ferroelectric, we have decided to study a well-known ferroelectric
BaTiO$_3$ doped with nickel, believing that its absorption spectrum can be
matched to the solar radiation spectrum.

Since nickel is used for fabricating contacts to ferroelectric ceramics,
at the first stages of studying nickel-doped barium titanate, the main attention
was focused on the determination of nickel solubility in BaTiO$_3$, its
influence on the Curie temperature, the dielectric constant, and other properties
as well as on the influence of Ni on the transition to a nonpolar hexagonal
$P6_3/mmc$ phase that is stable in undoped BaTiO$_3$ at temperatures above
1460$^\circ$C. It was shown that incorporation of Ni decreased the transition
temperature to the hexagonal phase; however, the nickel solubility and its
concentration needed for this transition were strongly dependent on synthesis
conditions and the available data are very contradictory.~\cite{ProcPhysSoc.76.763,
MaterChemPhys.105.320,JAlloysComp.481.559,SolidStateCommun.191.19,
JPhysCondensMatter.23.115903}  It was found that Ni in BaTiO$_3$ acts as an
acceptor.~\cite{SolidStateIonics.73.139} The doping of barium titanate with
nickel decreased the dielectric constant and the Curie temperature and also
resulted in a smearing of the ferroelectric phase transition in BaTiO$_3$ as the
Ni concentration increased.~\cite{JElectroceram.18.183,JElectroceram.21.394,
IndJEngMaterSci.16.390}  The kinetics of the transformation between the cubic
and hexagonal phases in BaTiO$_3$ doped with various impurities was studied
in Ref.~\onlinecite{ProcPhysSoc.76.763}. It was shown that nickel favors the
transition to the hexagonal phase, while the doping of BaTiO$_3$ with strontium
prevents this transition. In Ref.~\onlinecite{MaterChemPhys.105.320}, it was
noted that BaTiO$_3$ doped with nickel changes its color to dark-brown.

Information on the charge state and the structural position of Ni in BaTiO$_3$
was mainly obtained from the EPR data and is also very contradictory. The lines
observed in the EPR spectra were ascribed either to Ni$^+$ ions at the
$B$~sites~\cite{JPhysCondensMatter.19.496214} or to off-center Ni$^+$ ions at
the $A$~sites.~\cite{SolidStateCommun.116.133} In the EPR studies of nickel-doped
hexagonal BaTiO$_3$,~\cite{JPhysCondensMatter.23.115903} Ni$^{3+}$ ions
substituting for Ti$^{4+}$ ions in two different positions (Ti(1) and Ti(2))
were found. However, the concentration of these centers was less than 5\% of the
nominal amount of nickel; i.e., most of the Ni ions in the sample were in the
EPR-inactive state.

In Refs.~\onlinecite{BullRASPhys.80.1068,Ferroelectrics.501.1} we studied the
charge state of Ni in Ba$_x$Sr$_{1-x}$TiO$_3$ by XAFS (X-ray absorption fine
structure) spectroscopy and showed that it changed from 4 to 2.5 as the barium
content was increased. The first-principle calculations explained this result
by different energies of the oxygen vacancy formation in SrTiO$_3$ and BaTiO$_3$.
The calculations of the electronic structure of nickel-doped samples revealed
the formation of an impurity band in the forbidden gap of BaTiO$_3$ and
SrTiO$_3$ and explained the origin of the strong impurity absorption.

In Refs.~\onlinecite{SolidStateCommun.191.19,JMaterSci.51.10429}, the appearance
of the ferromagnetism in BaTiO$_3$(Ni) ceramic samples at 300~K was observed
and explained by the existence of oxygen vacancies~\cite{SolidStateCommun.191.19}
or the appearance of metallic nickel precipitates in the ceramics reduced in
an Ar+H$_2$ atmosphere.~\cite{JMaterSci.51.10429}  In addition, of interest
are the studies of BaTiO$_3$/Ni composites which exhibited interesting
multiferroic properties, in particular, the possibility of the magnetization
reversal by electric field,~\cite{ApplPhysLett.96.142509} and also the works
on development of a heterophase material with a very high dielectric constant
near the percolation threshold.~\cite{AdvMater.13.1541,Ferroelectrics.381.167}

From the literature review it follows that BaTiO$_3$(Ni) is indeed a promising
material that can provide an effective solar radiation absorption. Because of
this, the purpose of this work is to study the BaTiO$_3$(Ni) samples synthesized
in various conditions using X-ray diffraction and other structural methods.
Since the doping impurity can enter $A$ and $B$~sites of the perovskite
structure and exist in them in different charge states,~\cite{JETPLett.89.457}
we use the XAFS spectroscopy to determine the structural position, the local
environment, and the charge state of impurities in BaTiO$_3$.

\section{SAMPLES, EXPERIMENTAL AND CALCULATION TECHNIQUES}

Since the structure of BaTiO$_3$ can be transformed from the cubic to the hexagonal
as a result of doping with nickel, and the addition of strontium prevents this
transition,~\cite{ProcPhysSoc.76.763} 20\% SrTiO$_3$ was added to BaTiO$_3$ to
retain its cubic structure.

The Ba$_{0.8}$Sr$_{0.2}$TiO$_3$ samples doped with 3\% Ni were prepared by
solid-phase synthesis. The starting components were BaCO$_3$, SrCO$_3$,
nanocrystalline TiO$_2$ obtained by hydrolysis of tetrapropylorthotitanate and
dried at 500$^\circ$C, and Ni(CH$_3$COO)$_2$$\cdot$4H$_2$O. The components were
weighted in required proportions, ground in acetone, and calcined in air in
alumina crucibles at 1100$^\circ$C for 4--8~h. The prepared powders were ground
again and annealed in air at 1500$^\circ$C for 2~h. Then, a part of the samples
was additionally annealed at 1100$^\circ$C for 8~h, but, as was shown by the
measurements, this did not change their properties (the experiment was performed to
verify whether the nickel charge state would change as the annealing temperature
is changed). To incorporate the impurities into the $B$~site, the sample composition
was intentionally deviated from stoichiometry to excess Ba. The synthesis of
the NiTiO$_3$ and BaNiO$_{3-\delta}$ reference compounds was described in
Refs.~\onlinecite{JAdvDielectrics.3.1350031,PhysSolidState.56.449}. The phase
composition was checked by X-ray phase analysis.

The X-ray absorption spectra were measured in the regions of the extended fine
structure (EXAFS) and near-edge structure (XANES) on the KMC-2 station of the BESSY
synchrotron radiation source (the electron energy was 1.7~GeV and the maximum
current was 290~mA) at the Ni $K$-edge (8333~eV) at 300~K. The radiation was
monochromatized by a Si$_{1-x}$Ge$_x$(111) double-crystal monochromator. The
EXAFS spectra were recorded in fluorescence mode. The intensity $I_0$ of the
radiation incident on a sample was measured using an ionization chamber, and the
intensity of the fluorescence radiation $I_f$ was measured using a R{\"O}NTEC
X-flash silicon energy-dispersive detector with an operating area of 10~mm$^2$.

The EXAFS spectra were analyzed using the widely used \texttt{IFEFFIT} program
package.~\cite{IFEFFIT}  An oscillating EXAFS function $\chi(k)$ was extracted
from the fluorescence excitation spectra $\mu(E) = I_f/I_0$ (where $E$ is the
X-ray quantum energy) using the \texttt{ATHENA} program. Then, by minimizing
the root-mean-square deviation of the experimental $k^2 \chi(k)$ curve from the
curve calculated for a given model of the local environment of the impurity atom
using the \texttt{ARTEMIS} program, the distances $R_j$ and the Debye--Waller
factor $\sigma^2_j$ for the $j$th shell were calculated. Simultaneously, the
energy origin correction $dE_0$ was also varied. The coordination numbers were
taken to be constant and were given by the structural model. The number of the
adjustable parameters (9--15) was smaller than the number of independent data
points $N_{\rm ind} = 2 \Delta k \Delta R/\pi \approx 20$. The dependences of
the scattering amplitude and phase, the phase shift of the central atom, and
the photoelectron free path on the photoelectron wave vector $k$ for all
single- and multiple-scattering paths required to build theoretical curves
$\chi(k)$ were calculated using the \texttt{FEFF6} program.~\cite{FEFF}  An
important feature of the \texttt{ARTEMIS} program is the possibility of
fitting of the EXAFS spectra to a sum of several independent contributions.

An important feature of this work is that the pure geometric approach characteristic
of standard EXAFS data analysis is complemented with the calculations of the
geometry and the electronic structure of impurity complexes, which enables us
to estimate the physical feasibility of different structural models, and so to
increase the reliability of the results.

The geometry and the electronic structure of nickel-doped BaTiO$_3$ were modeled
from first principles within the density functional theory using the \texttt{ABINIT}
software package. The calculations were performed on 40-atom (simple cubic) and
80-atom (fcc) supercells in which one of the Ti atoms was replaced by a Ni atom
(the Ni concentrations were 12.5 and 6.25\%, respectively). Since the nickel
atom has a partially filled $d$~shell, we performed the calculations using PAW
pseudopotentials~\cite{ComputMaterSci.81.446} and the LDA+$U$
approximation.~\cite{JPhysCondensMatter.9.767}  Parameters $U = 5$~eV
and $J = 0.9$~eV that describe the Coulomb and the exchange interactions inside
the $d$ shell were taken from the literature as typical values of these parameters
for Ni. It was shown that a variation of these parameters within 20\% did not
substantially influence the results. The cutoff energy was 30~Ha (816~eV); the
integration over the Brillouin zone was performed on a 4$\times$4$\times$4
Monkhorst--Pack mesh for the simple cubic supercell or the mesh with equivalent
$k$-point density for the fcc supercell. The relaxation of the lattice parameters
and the atomic positions in the supercells was stopped when the Hellmann--Feynman
forces became less than 10$^{-5}$~Ha/Bohr (0.5 meV/{\AA}).

The modeling on 40-atom and 80-atom supercells without oxygen vacancies gives
the results for the Ni$^{4+}$ charge state of the impurity. In order to change
the charge state to Ni$^{2+}$, either a vacancy was introduced into the system
or we used a trick~\cite{Ferroelectrics.206.69} in which two extra electrons
were added to the system. Although, in the latter case, the system was not
electrically neutral, the tests showed that the addition of two electrons to the
80-atom supercell of barium titanate results in a small increase in the energy
gap width (0.12~eV) and in the lattice parameter (0.61\%). However, the obtained
density of states and the Ni--O interatomic distances in doped samples were close
to those calculated for a model in which the Ni$^{2+}$ charge state was
obtained by adding a distant oxygen vacancy located at distance of 5.8~{\AA}
from the Ni atom.

\section{EXPERIMENTAL RESULTS}

\subsection{X-Ray diffraction measurements}

\begin{figure}
\includegraphics{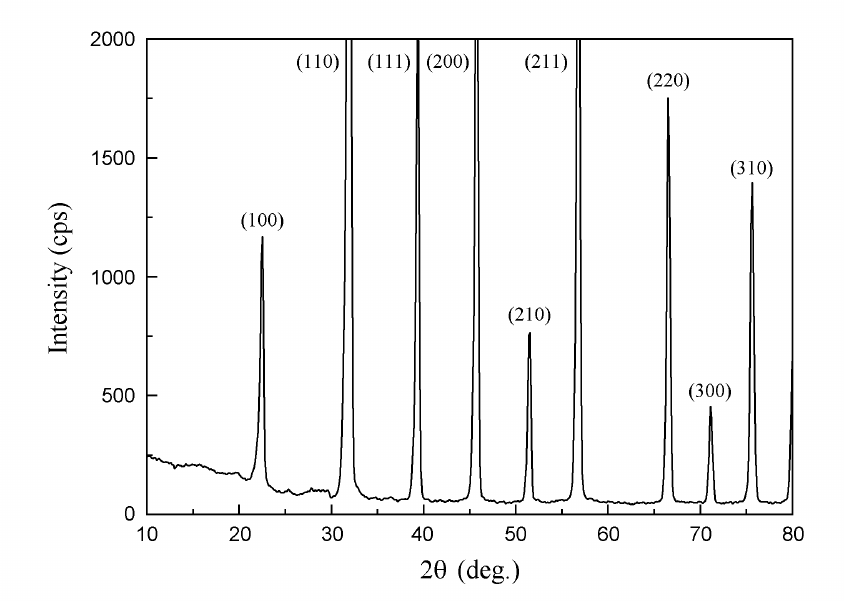}
\caption{\label{fig1}X-ray diffraction pattern of the
Ba$_{0.8}$Sr$_{0.2}$TiO$_3$(3\% Ni) sample.}
\end{figure}

Fig.~\ref{fig1} shows a typical X-ray diffraction pattern of the samples under
study. The Ba$_{0.8}$Sr$_{0.2}$TiO$_3$ samples doped with 3\% Ni have the
perovskite-type cubic structure with the lattice parameter $3.9723 \pm 0.0013$~{\AA}
at 300~K and were almost single-phase. Small traces of a second phase in the
angular range $2\theta = {}$20--30$^\circ$ belong to the Ba$_2$TiO$_4$ compound
that often appears in ceramic barium titanate. Comparing the obtained lattice
parameter with the lattice parameter of undoped Ba$_{0.8}$Sr$_{0.2}$TiO$_3$
solid solution ($a = 3.978$~{\AA}) shows a slight compression of the lattice
which indicates that the impurity incorporated into the sample. The above lattice
parameter for the undoped solid solution was obtained by an extrapolation of the
data of Ref.~\onlinecite{JAmCeramSoc.38.444} from the cubic phase to the region,
in which the solid solution became tetragonal as a result of the ferroelectric
phase transition. The incorporation of nickel to the solid solution decreased the
Curie temperature in our samples below 300~K.

\subsection{Analysis of the XANES spectra}
\label{sec32}

\begin{figure}
\includegraphics{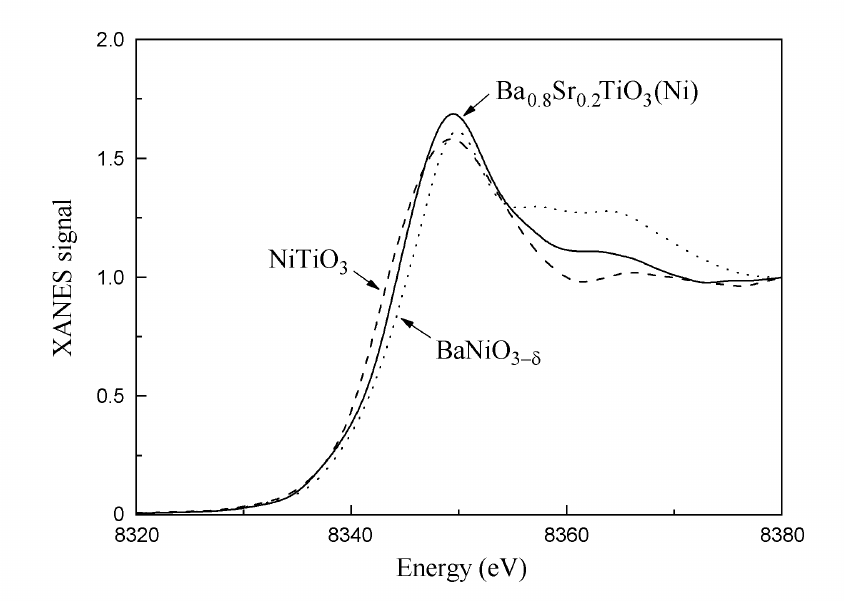}
\caption{\label{fig2}XANES spectra of the Ba$_{0.8}$Sr$_{0.2}$TiO$_3$(3\% Ni)
sample and the reference nickel compounds.}
\end{figure}

To determine the charge state of the Ni impurity in Ba$_{0.8}$Sr$_{0.2}$TiO$_3$,
we compared the position of the absorption edge in the XANES spectra of the
samples under study with the positions of the edges in the BaNiO$_{3-\delta}$
and NiTiO$_3$ reference compounds (Fig.~\ref{fig2}). It is seen that the
absorption edge in the Ba$_{0.8}$Sr$_{0.2}$TiO$_3$(3\% Ni) sample is shifted
by 0.7~eV from the absorption edge in NiTiO$_3$, where the nickel charge state
is 2+, to the absorption edge in the BaNiO$_{3-\delta}$ reference compound in
which the nickel charge state is 3.4+ (the nickel charge state in this sample
was determined in Refs.~\onlinecite{JAdvDielectrics.3.1350031,PhysSolidState.56.449}).
From the obtained absorption edge shift (see the discussion in
Refs.~\onlinecite{BullRASPhys.80.1068,Ferroelectrics.501.1}) it was concluded
that the mean charge state of Ni in Ba$_{0.8}$Sr$_{0.2}$TiO$_3$(3\% Ni) sample
is $\sim$2.5+. This implies that the most part of the Ni ions in the samples
is in charge state 2+ and only small part of them is in charge states 3+ or 4+.
Our conclusion agrees with the results of the X-ray photoelectron spectroscopy
study~\cite{SolidStateCommun.191.19} in which the main charge state of Ni in
BaTiO$_3$ was 2+.

\subsection{Analysis of the EXAFS spectra}
\label{sec33}

To determine the structural position and the local environment of the Ni impurity
in Ba$_{0.8}$Sr$_{0.2}$TiO$_3$, the EXAFS spectra were analyzed. The models of
impurity centers were constructed taking into account the charge state of the
impurity obtained from the XANES spectra. First, we analyzed the models
in which the impurity atom entered either $A$ or $B$~sites of the perovskite
structure with a different local environment. A criterion of agreement between
the experimental and calculated spectra was a small value of the $R$~factor
that quantitatively describes the discrepancy of the compared curves, the
closeness of the value of the $S_0^2$ parameter to the value obtained from the
analysis of the EXAFS spectra of the reference compounds, and a visual agreement
of the Fourier-transforms of the spectra in the $R$~representation.

The EXAFS data analysis of the NiO, NiTiO$_3$, and metallic Ni reference compounds
enabled us to determine the $S_0^2$ value, which was equal to 0.92 for NiTiO$_3$,
0.97 for NiO, and 0.77 for Ni.

An analysis of a model in which the Ni$^{2+}$ ion enters the $A$~site both in
the on-center and off-center positions showed that the calculated spectrum
on no account can describe well the experimental spectrum (the least value of
the $R$~factor was 0.408).

For an impurity atom at the $B$~site (Ni$^{2+}$ substitutes for Ti$^{4+}$), the
sample must contain vacancies to provide the electrical neutrality. Thus, we
considered two models: the model with an oxygen vacancy $V_{\rm O}$ located in
the first shell of the Ni atom (in this case, the shift of the Ni atom from the
site was allowed to be up to 0.5~{\AA} in the direction of the vacancy) and the
model with a distant oxygen vacancy and the on-center position of the impurity.

\begin{table}
\caption{\label{table1}Local structure of the Ba$_{0.8}$Sr$_{0.2}$TiO$_3$(3\% Ni)
sample (data analysis within a single model).}
\begin{ruledtabular}
\begin{tabular}{cccc}
Adjustable & Site $A$ & Ni$^{2+}_B$--$V_{\rm O}$ & Ni$^{2+}_B$--$V_{\rm O}$ \\ 
parameter  &          & (distant vacancy)        & (nearest vacancy) \\ 
\hline
$R$~factor & 0.408 & 0.0741 & 0.0537 \\
$S_0^2$    & 0.896 & 0.774  & 0.889 \\
$R_{\rm Ni-O(I)}$ ({\AA}) & 2.086 & 2.080 & 2.070 \\
$R_{\rm Ni-O(II)}$ ({\AA}) & 3.120, 3.678 & --- & 1.873 \\
$R_{\rm Ni-Ba}$ ({\AA})  & 2.752 & 3.436 & 3.383--3.499 \\
$R_{\rm Ni-Ti}$ ({\AA})  & 2.974 & 3.918 & 3.786--3.986 \\
\end{tabular}
\end{ruledtabular}
\end{table}

A reasonable agreement between the experimental and calculated spectra was
obtained in two cases, in which (1) Ni$^{2+}$ atom substitutes for the Ti$^{4+}$
atom and there is a distant oxygen vacancy $V_{\rm O}$ and (2) Ni$^{2+}$ atom
substitutes for the Ti$^{4+}$ atom and is displaced from the $B$~site toward the
vacancy $V_{\rm O}$ located in the first shell. Table~\ref{table1} gives the
interatomic distances to the nearest shells and the values of the $R$~factor
obtained for these models. It is seen that in the model with a distant vacancy,
the Ni--O distances are close to the sum of the ionic radii of the O$^{2-}$ and
Ni$^{2+}$ ions ($R_{\rm Ni^{2+}} = 0.69$~{\AA}, $R_{\rm O^{2-}} = 1.4$~{\AA}),
so that the XANES and EXAFS data agree well. However, the $S_0^2$ parameter in
this model is underestimated. As for the second model, unfortunately, we did not
get a single-valued solution. In addition to the solution reported in
Refs.~\onlinecite{BullRASPhys.80.1068,Ferroelectrics.501.1},
we found two more possible solutions: one, in which the Ni atom is displaced by
0.1~{\AA} toward the vacancy, and the other, in which the Ni atom is displaced
in the opposite direction. Table~\ref{table1} gives the data for the latter
solution which had the lowest $R$~factor. The appearance of several solutions
with close $R_{\rm Ni-O(I)}$ distances to four atoms and substantially different
$R_{\rm Ni-O(II)}$ distances to the fifth atom (2.438, 2.178, and 1.873~{\AA})
is related to the fact that the contribution of a single O(II) atom to the
EXAFS function is small, and the nonlinear regression problem can find one of
numerous solutions corresponding to a local, not global, minimum.

\section{Results of first-principle calculations}
\label{sec4}

The analysis of the experimental data in Sec.~\ref{sec33} enabled us to propose
several structural models that reasonably describe the experimental EXAFS
spectrum. To choose the most adequate model, we used the first-principle
calculations to model the geometry of a number of impurity complexes containing
nickel. In the modeling, particular attention was paid to configurations in
which nickel had the charge state 2+, which is closest to that obtained in the
experiment.

Although we have not obtained evidences that the nickel atoms enter the $A$~sites
of the perovskite structure, we considered this case, too. The calculations
showed that the on-center position of the Ni$^{2+}$ ion at the $A$~site is
unstable. The modeling for an off-center Ni$^{2+}$ ion at the $A$~site showed
that this impurity center is diamagnetic and is characterized by nearly square
planar configuration (the shift of Ni from the plane of four oxygen atoms is
0.195~{\AA}) and short Ni--O distances (1.867~{\AA}) that do not agree with
the experiment.

In the case when a Ni atom enter the $B$~site, the simplest case is that
in which the nickel charge state coincides with the titanium one (4+). In
this case, the electrical neutrality is not violated upon doping, and
the defect can be considered as a point defect. The calculations of the
geometry of the 80-atom supercell showed that, as a result of relaxation,
the distances to the oxygen atoms in the first shell are 1.922~{\AA}. The
ion is in the diamagnetic state ($S = 0$). The comparison of the calculated
distances with those obtained from the EXAFS data analysis shows their marked
difference. So, the geometry of this defect does not agree with that observed
in the experiment.

Actually, the local environment of nickel in barium titanate is more
complicated, since the impurity charge (2+) found in our experiment suggests
that the structure should contain an oxygen vacancy that compensates the
difference of the charges of the Ni$^{2+}$ and Ti$^{4+}$ ions.

In our calculations, we first assumed that a vacancy in the formed
Ni$^{2+}$--$V_{\rm O}$ complex is located in the first shell (this structure
completely corresponds to the model considered in Ref.~\onlinecite{PhysRevB.83.205115}).
The calculations of the equilibrium configuration of this complex performed
on a 80-atom supercell showed that the Ni ion is shifted by only 0.021~{\AA}
from the plane formed by four oxygen atoms. The distance to four nearest O atoms
in the plane is 1.925~{\AA} and the distance to the fifth oxygen atom is
2.192~{\AA}. The impurity complex is diamagnetic ($S = 0$). A comparison of
the calculated interatomic distances with those obtained from the EXAFS data
analysis shows that this model also does not agree with the experiment.

\begin{table}
\caption{\label{table2}Magnetic state and local environment of Ni atoms
in a number of compounds of two-, tri-, and tetravalent nickel.}
\begin{ruledtabular}
\begin{tabular}{cccc}
Compound & Charge & Magnetic & $R_{\rm Ni-O}$ ({\AA}) \\
         & state  & state    & \\
\hline
NiO ($R{\bar 3}m$) & 2+ & $S = 1$ & 2.044 \\
NiTiO$_3$ ($R{\bar 3}$) & 2+ & $S = 1$ & 1.993--2.034 \\
NiNb$_2$O$_6$ ($Pbcn$) & 2+ & $S = 1$ & 1.983--2.014 \\
BaNiO$_2$ ($Cmcm$) & 2+ & $S = 0$ & 1.850 \\
LiNiO$_2$ ($R{\bar 3}m$) & 3+ & $S = 1/2$ & 1.926 \\
BaNiO$_3$ ($P6_3/mmc$) & 4+ & $S = 0$ & 1.849 \\
SrNiO$_3$ ($P6_3/mmc$) & 4+ & $S = 0$ & 1.842 \\
PbNiO$_3$ ($R3c$) & 4+ & $S = 1$ & 1.974--2.026 \\
\end{tabular}
\end{ruledtabular}
\end{table}

Prior to further discussion, we should make an important remark. In all
aforementioned models, the distances to the nearest oxygen atoms were close
to each other and were slightly dependent on the ion charge. This put us on
the idea to perform additional calculations for a number of compounds of
two-, tri-, and tetravalent nickel (NiO, NiTiO$_3$, NiNb$_2$O$_6$, BaNiO$_2$,
BaNiO$_3$, LiNiO$_2$, SrNiO$_3$, and PbNiO$_3$). The calculation showed that
a ``long'' distance (1.97--2.04~{\AA}) is obtained only when the nickel atom
is paramagnetic ($S = 1$) (Table~\ref{table2}). As follows from
Table~\ref{table2}, the interatomic Ni--O distance is determined primarily by
the spin state of the Ni atom rather than by its charge, and so we first should
focus our attention on the paramagnetic centers to explain the geometry of an
impurity Ni center in BaTiO$_3$.

It turned out that the complexes of the Ni$^{2+}$ ion with the oxygen vacancy
located in the fourth or more distant shell of the Ni atom (``a distant vacancy'')
are such centers. We considered three different configurations of the complexes.
In the first of them, the vacancy was located at a distance of 1.5~lattice
constants along the [001] axis; in the second of them, the vacancy was
additionally displaced sideways. In the third configuration, two extra electrons
were added to the system (these electrons can be supplied by either oxygen
vacancies or donor impurity atoms). The relaxation of the positions of all atoms
in these models showed that the mean Ni--O distances in them are 2.038--2.082~{\AA},
in better agreement with the experiment.

\section{Discussion}

As follows from Sec.~\ref{sec4}, the interatomic Ni--O distance is determined by
the magnetic state of the Ni ion to a greater extent than by its charge state
(Table~\ref{table2}). In addition, it was shown that the Ni$^{2+}$--$V_{\rm O}$
complex with the nearest vacancy considered in Ref.~\onlinecite{PhysRevB.83.205115}
is diamagnetic and is
characterized by a set of distances that strongly differ from those obtained
from the EXAFS data analysis. The Ni--O distance that was observed in the
experiment, in our opinion, can be explained by Ni$^{2+}$ paramagnetic
complexes with distant vacancies. However, the analysis performed in Sec.~\ref{sec33}
showed that in this case we are dealing with an underestimated value of $S_0^2$
(Table~\ref{table1}).

To solve this problem, we considered a number of new structural models. First,
we assumed that a part of the Ni atoms in Ba$_{0.8}$Sr$_{0.2}$TiO$_3$ is in
a higher charge state, which follows from the analysis of the XANES spectra
(Sec.~\ref{sec32}). We considered a model, in which the nickel atom is at the
$B$~site in two charge states simultaneously. We took into account a possible
difference in the distances to the nearest oxygen atoms and the difference in
the absorption edge energies of the atoms in different charge states. A similar
situation can be also observed in a model in which the Ni atoms have the same
charge state but different magnetic states ($S = 0$ and 1); such a situation is
possible, for example, when the Ni$^{2+}$ ions form the complexes with the
neighboring and distant vacancies simultaneously. The difference of this model
from the previous one is only in that the absorption edge energies are
considered to be the same in the latter case.

\begin{table*}
\caption{\label{table3}Local structure of the Ba$_{0.8}$Sr$_{0.2}$TiO$_3$(3\% Ni) sample
(fitting with the determination of two sets of structural parameters simultaneously)}
\begin{ruledtabular}
\begin{tabular}{ccccc}
Adjustable & \multicolumn{4}{c}{Model} \\
\cline{2-5}
parameter  & Two spin states, & Two charge states,       & Ni$^{2+}$ at the $B$~site and the  & Ni$^{2+}$ at the $B$~site and the \\
           & Ni$^{2+}$        & Ni$^{4+}$ and Ni$^{2+}$  & hexagonal phase precipitates       & second BaNiO$_{3-\delta}$ phase \\
\hline
$R$~factor & 0.0627 & 0.0618 & 0.0288 & 0.00147 \\
\multicolumn{5}{c}{The first set of parameters} \\
& $S = 0$  & Ni$^{4+}$ & Cubic phase & Ni$^{2+}$ \\
$S_0^2$    & 0.472 & 0.546 & 0.893 & 0.465 \\
$R_{\rm Ni-O(I)}$ ({\AA}) & 1.917 & 1.906 & 2.123 & 2.113 \\
$R_{\rm Ni-Ba}$ ({\AA}) & 3.434 & 3.435 & 3.440 & 3.456 \\
$R_{\rm Ni-Ti}$ ({\AA}) & 3.915 & 3.911 & 3.972 & 3.972 \\
\multicolumn{5}{c}{The second set of parameters} \\
& $S = 1$  & Ni$^{2+}$ & Hexagonal phase & Phase BaNiO$_{3-\delta}$ \\
$S_0^2$    & 0.593 & 0.781 & 0.165 & 0.377 \\
$R_{\rm Ni-O(I)}$ ({\AA}) & 2.097 & 2.088 & 2.064 & 1.899 \\
$R_{\rm Ni-Ba}$ ({\AA}) & 3.434 & 3.435 & 3.465--3.574 & 2.416 (Ni--Ni)$^*$ \\
$R_{\rm Ni-Ti}$ ({\AA}) & 3.915 & 3.911 & 3.919 & 3.512 (Ni--Ba)$^*$ \\
\end{tabular}
\end{ruledtabular}
{\footnotesize $^*$ The interatomic distances and corresponding atoms in the BaNiO$_3$ structure. \hfill}
\end{table*}

The results obtained from the analysis of these models are given in
Table~\ref{table3}. Although these models describe the experimental $k^2 \chi(k)$
curve quite reasonably, we again encounter an underestimated value of $S_0^2$.

Lastly, the last group of the models were the models that admit the existence
of the second phase precipitates in the samples. The first of them was
a model in which the cubic Ba$_{0.8}$Sr$_{0.2}$TiO$_3$(3\% Ni) phase and
microprecipitates of the hexagonal BaTiO$_3$(Ni) phase coexist. Both variants
of incorporation of nickel into the hexagonal phase (Ni at the Ti(1) and Ti(2)
sites) were analyzed. Although these models gave a reasonable agreement for
a sum of $S_0^2$ parameters of both the phases, the values of the $R$~factor
characterizing the agreement between the curves were not quite low. The
parameters for the model with nickel at site Ti(1) are given in Table~\ref{table3}.

Then we analyzed the models in which the samples were supposed to contain
microinclusions which could appear at the grain boundaries in our ceramic
powders and which are hardly detected using X-ray diffraction (X-ray amorphous
phase). We considered the variants of inclusions of NiTiO$_3$, NiO, metallic Ni,
and the BaNiO$_{3-\delta}$ compound. An analysis of the EXAFS spectra showed
that the variants with inclusions of NiTiO$_3$, NiO, and metallic Ni very
insignificantly improved the agreement between the calculated and experimental
EXAFS spectra since the fraction of the second phase was from 0.001 to 0.017,
i.e., it was close to zero (the parameters describing the local structure of
the second phase were taken from the EXAFS data analysis of pure individual
phases). In these models the $R$~factor was about 0.066.

\begin{figure}
\includegraphics{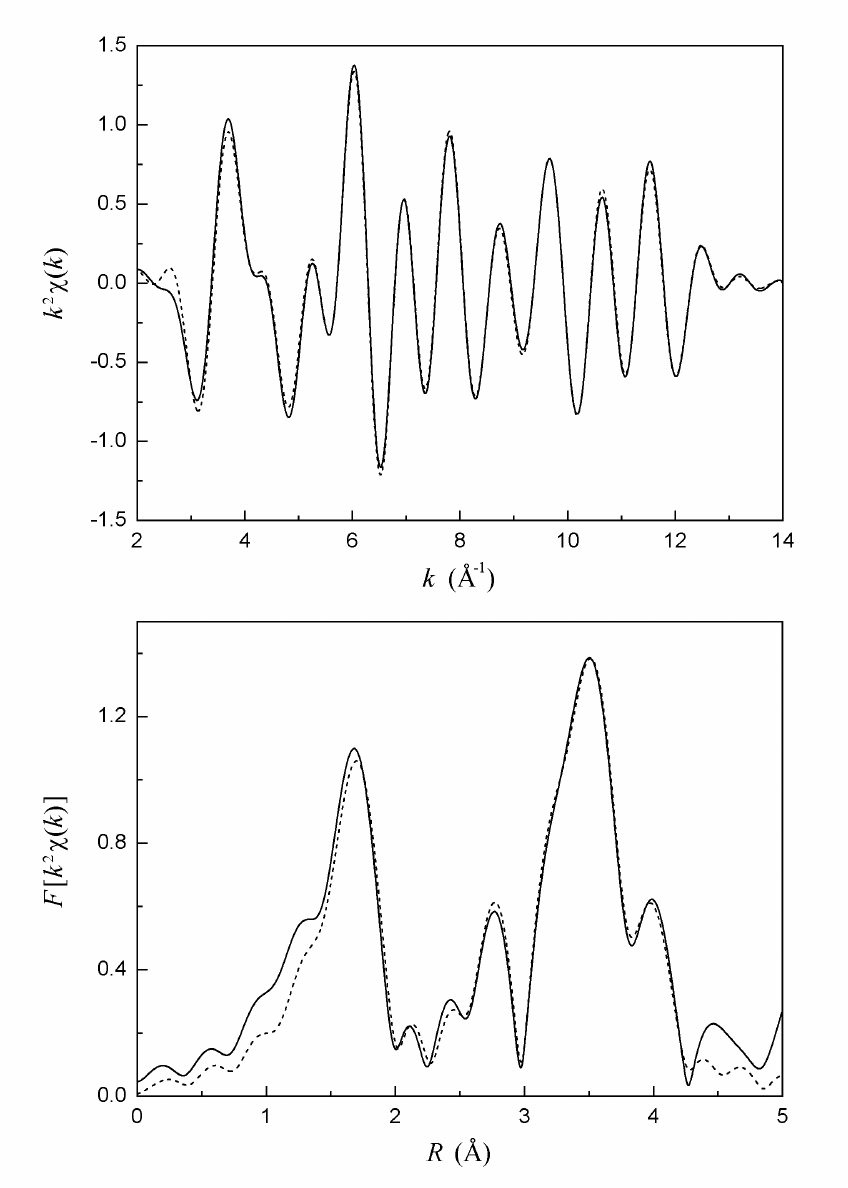}
\caption{\label{fig3}EXAFS spectra of the Ba$_{0.8}$Sr$_{0.2}$TiO$_3$(3\% Ni)
sample in (a) $k$- and (b) $R$-representations: the solid lines show the
experimental data, the dashed lines are the result of fitting
performed in the range $k = {}$3--12.4~{\AA}$^{-1}$ and $R = {}$1--4.3~{\AA}.}
\end{figure}

The best agreement between the calculated and experimental spectra was
obtained for a model in which the second phase is BaNiO$_{3-\delta}$. For
this model, the $R$~factor decreased by more than one order of magnitude as
compared to that for the models considered above (Table~\ref{table3}).
Fig.~\ref{fig3} shows a typical EXAFS spectrum $k^2 \chi(k)$ for
Ba$_{0.8}$Sr$_{0.2}$TiO$_3$(3\% Ni) and its Fourier transform $F(R)$, which
are compared to the results of calculations within this model. Unfortunately,
a strongly defective structure of BaNiO$_{3-\delta}$, whose composition
($0 \le \delta \le 1$) depends on the temperature and partial oxygen pressure
during its synthesis, made it impossible to use the experimental EXAFS spectra
in the analysis. Because of this, the structure of BaNiO$_{3-\delta}$ was
modeled from first principles using the structural model of BaNiO$_3$. The
results of the fitting are given in Table~\ref{table3}. The most important in
this model is that the value of $S_0^2$ finally became close to the values
obtained for the reference compounds. Taking into account that the second
phase is formed by an individual nickel compound and the main phase is the
solid solution with a nickel concentration lower than 3\%, the ratio of the
volumes of these phases can be estimated as 1:50, based on the ratio of
contributions of the two phases (values of $S_0^2$). Note that the
distance of 2.438~{\AA}, which was earlier attributed to the Ni--O
one,~\cite{JAdvDielectrics.3.1350031,PhysSolidState.56.449}  actually
corresponds to the Ni--Ni distance in BaNiO$_{3-\delta}$.

Note that two pure compounds in the BaNiO$_{3-\delta}$ system, namely
BaNiO$_2$ and BaNiO$_3$, have similar local structures. They consist of
one-dimensional Ni--Ni chains extending along one of the axes with quite
short Ni--Ni distance (about 2.4~{\AA}) and two or three oxygen atoms located
adjacent to this short bond. When there are two atoms, they are arranged as
planar NiO$_4$ squares forming puckered chains; when there are three atoms,
they form triangles whose centers are in the middle of the Ni--Ni bond. When
the number of oxygen atoms is intermediate, as is the case of BaNiO$_{3-\delta}$,
the atoms are located in random positions adjacent to the Ni--Ni chains.
There is no a long-range order in the arrangement of these atoms, and they
are highly disordered. This explains very high values of the Debye--Waller
factors for the oxygen atoms in the second phase and a practical impossibility
of observing precipitates of this phase in X-ray diffraction. To simulate
the defective BaNiO$_{3-\delta}$ phase, we chosen the BaNiO$_3$ phase,
although either of two phases under discussion can be used when analyzing
the EXAFS spectra.

In conclusion, we discuss the decrease in the lattice parameter observed in
our samples upon their doping. According to our measurements, the lattice
parameter of the nickel-doped samples was 0.14\% less than that of the undoped
material. Our calculations of the changes in the cubic lattice parameter upon
doping gave the following results for the nickel concentration of 6.25\%:
+0.522\% for the complex of Ni$^{2+}$ with a distant vacancy, $-$0.095\%
for the complex of Ni$^{2+}$ with the nearest vacancy, and $-$0.74\% for
Ni$^{4+}$ at the $B$~site. It is seen that the model of the complex with
a distant vacancy proposed in this work do not agree well with the X-ray
diffraction data. However, it should be noted that the lattice parameter also
decreased in a nickel-doped SrTiO$_3$, in which nickel entered the $B$~site as
a Ni$^{4+}$ ion. Taking into account that the mean charge state in our samples
is $\sim$2.5+, we can suppose that a part of nickel in the samples has the
charge state 4+. The systematic appearance of the solutions with shorter Ni--O
distances shows that such short distances can be in the samples (Table~\ref{table3}).
Unfortunately, the fitting of the EXAFS spectra with simultaneous account of
three models of the local environment is impossible because of a very large
number of adjustable parameters.

\section{Conclusions}

The local environment and the charge state of the nickel impurity in cubic
Ba$_{0.8}$Sr$_{0.2}$TiO$_3$ have been studied by XAFS spectroscopy. The analysis
of the XANES spectra showed that the mean charge state of the Ni atoms is
$\sim$2.5+. The analysis of the EXAFS spectra, which took into account the
results of first-principle calculations of the geometry of defects, suggests
that Ni atoms enter the $B$~sites of the perovskite structure, and the difference
between the Ni$^{2+}$ and Ti$^{4+}$ charges is mainly compensated by distant
oxygen vacancies. In addition, the EXAFS data analysis revealed that a
considerable amount of nickel in the sample is in the form of the X-ray
amorphous second phase BaNiO$_{3-\delta}$. The measurement of the lattice
parameter and its comparison with the results of the computer modeling
showed that the decrease in the unit cell volume upon doping could indicate
the existence of a small amount of Ni$^{4+}$ ions at the $B$~sites in the
samples. The obtained results can be interesting for designing a new type of
solar energy converters that use the bulk photovoltaic effect in the material
studied in this work.

\begin{acknowledgments}
The authors are grateful to the BESSY staff for the possibility of performing
the experiments and technical assistance.

This work was supported by the Russian Foundation for Basic Research, project
no. 17-02-01068.
\end{acknowledgments}

\providecommand{\BIBYu}{Yu}

\end{document}